\documentclass[aps,twocolumn,groupedaddress,showpacs]{revtex4}
\usepackage{graphicx}

\begin{document}

\title[Latent Heat Calculation of the 3D q-state Potts models by TPVA]
{Latent Heat Calculation of the 3D q=3, 4, and 5 Potts models by Tensor
Product Variational Approach}

\author{A.~Gendiar}
\email{gendiar@savba.sk}
\affiliation{Institute of Electrical Engineering, Slovak Academy of Sciences,
D\'{u}bravsk\'{a} cesta 9, SK-842 39 Bratislava, Slovakia}
\affiliation{Department of Physics, Faculty of Science,
Kobe University, 657-8501, Japan}
\author{T.~Nishino}
\email{nishino@phys.sci.kobe-u.ac.jp}
\affiliation{Department of Physics, Faculty of Science,
Kobe University, 657-8501, Japan}

\date{\today}

\begin{abstract}
Three-dimensional (3D) $q$-state Potts models ($q$=3, 4, and 5) are studied by
the tensor product variational approach (TPVA), which is a recently developed
variational method for 3D classical lattice models. The variational state is given by
a two-dimensional (2D) product of local factors, and is improved by way of
self-consistent calculations assisted by the corner transfer matrix renormalization
group (CTMRG). It should be noted that no a priori condition is imposed
for the local factor. Transition temperatures and latent heats are calculated from
the observations of thermodynamic functions in both ordered and disordered phases.
\end{abstract}


\maketitle

\section{Introduction}

The density matrix renormalization group (DMRG), which was invented by
White in 1992~\cite{DMRG1,DMRG2}, has been applied to a wide class of
one-dimensional (1D) quantum systems including quantum spin ladders~\cite{DMR}.
DMRG is also efficient for obtaining thermodynamic functions of two-dimensional
(2D) classical systems~\cite{Nishi,DMR}. Now a technical interest in DMRG is to
extend its applicability to higher dimensional systems~\cite{Liang,Xiang,Peschel}.

It is worth looking at the variational background in DMRG in order to obtain a
rough image of DMRG in higher dimension.  In 1995 Ostlund
and Rommer showed that DMRG assumes so called `the matrix product
wave function', and that a very small numbers of parameters
are sufficient to obtain a good variational energy~\cite{Ostlund}.
It is a small surprise that such a construction of variational state has
been known for long years in the field of statistical mechanics of 2D
classical lattice models. In 1945 Kramers and Wannier introduced a
very simple matrix product as a variational state for the transfer
matrix of the 2D Ising model~\cite{KWA}. Later, the idea of constructing variational
state from local elements was extended by Kikuchi~\cite{Kikuchi}
(the cluster approximation), Baxter~\cite{Bax1,Bax2},
and Villani~\cite{Villani} (the correlation length equality approach).
All these approaches calculate the lower bounds of the
free energies of a 2D system. They use a variational state that corresponds to
an effective 1D statistical system with several adjustable parameters.

Simply increasing the space dimension by one, we can extend such
variational formula to 3 dimensions. The simplest example is the
Kramers-Wannier approximation applied to the 3D Ising model
by Okunishi and Nishino~\cite{KW_ON},
where the 2D Ising model under the external magnetic field is treated as
variational state, which has only two adjustable parameters.
The calculated spontaneous magnetization and transition temperature
are more precise than those obtained from a former attempt
to extend DMRG to 3D classical systems~\cite{CTTRG}. A major problem in 
the Kramers-Wannier approximation is 
that one can not always find out a good functional form of variational 
state intuitively, esp. for models other than the
3D Ising model. In order to overcome this problem, a
numerical self-consistent approach has been introduced,
which we call `the tensor product variational approach (TPVA)' in the
following~\cite{TPVA1,TPVA2}. In TPVA the variational state is
determined automatically, with no reference to a priori information on
systems. In this paper we briefly review the variational principle and the numerical
algorithm of TPVA, and discuss the applicability of this method via trial
calculations for $q=3, 4, 5$ Potts models.

In Sec.~II we introduce main features of the new
algorithm from the variational point of view. We focus on the 
self-consistent improvement of the variational state. A specific way how 
to apply the variational method to the Potts model is presented in Sec.~III.
We also provide the way how to calculate the internal 
energy and the magnetization. The numerical results are presented in Sec.~IV.
In Sec.~V we conclude the main results.

\section{Variational approach in two dimensions}

For a tutorial purpose we first explain the way how to apply TPVA to the square 
lattice Potts model. (Later in the next section we treat the cubic lattice.)

\begin{figure}[!ht]
\includegraphics[width=70mm,clip]{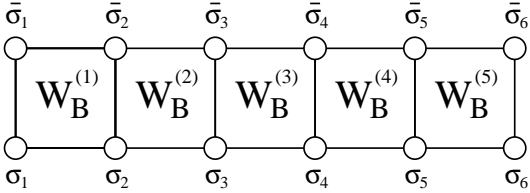}
\caption{The transfer matrix ${\cal T}[ \bar\sigma | \sigma ]$ in Eq.~(\ref{T2D}). 
This is the
case where $2N = 6$ and therefore there are five Boltzmann weights $W_{\rm B}^{(i)}$
from $i = 1$ to $i = 5$.}
\label{TM2D}
\end{figure}

Let us consider an infinitely long stripe of the width $2N$ on the square lattice,
which is nothing but the $2N$-leg ladder, and consider the $q$-state Potts model 
in this finite width region. Fig.~\ref{TM2D} shows the transfer matrix
${\cal T}[ \bar\sigma | \sigma ]$ of this system when $2N = 6$, where
\begin{equation}
[ \sigma ] =
( \sigma_1^{~}, \sigma_2^{~}, \ldots, \sigma_{2N}^{~} )
\,\,\, \mbox{and} \,\,\,
[ \bar\sigma ] =
( \bar\sigma_1^{~}, \bar\sigma_2^{~}, \ldots, \bar\sigma_{2N}^{~} )
\label{spin}
\end{equation}
represent adjacent rows of $q$-state spin variables.
Here we interpret the Potts model as a special case of so called `the interaction
round a face' (IRF) model~\cite{Baxter}, and construct 
${\cal T}[ \bar\sigma | \sigma ]$ as a product of plaquette Boltzmann weights
\begin{equation}
{\cal T}[ \bar\sigma | \sigma ] = \prod_{i=1}^{2N-1} W_{\rm B} 
(\bar\sigma_i^{~} \bar\sigma_{i+1}^{~} | \sigma_i^{~} \sigma_{i+1}^{~})=
\prod_{i=1}^{2N-1}W_{\rm B}^{(i)}\{ \bar\sigma | \sigma \} \, ,
\label{T2D}
\end{equation}
where we have written the nearest neighbor spin pairs
$( \sigma_i^{~} \sigma_{i+1}^{~} )$ and
$( \bar\sigma_i^{~} \bar\sigma_{i+1}^{~} )$, respectively, as
$\{ \sigma \}$ and $\{ \bar\sigma \}$ for the book keeping purpose.
Following this index rule, the local Boltzmann weight is written as follows
\begin{eqnarray}
&&
W_{\rm B}^{(i)}\{ \bar\sigma | \sigma \} =
W_{\rm B}^{~}( \bar\sigma_{i} \bar\sigma_{i+1} | \sigma_{i} \sigma_{i+1} )\\
=&&\hspace{-0.2cm}
\exp\left[ -\frac{J}{2k_{\rm B} T}
\left(
\delta_{\sigma_{i}\sigma_{i+1}}+\delta_{\bar\sigma_{i}\bar\sigma_{i+1}}+
\delta_{\sigma_{i}\bar\sigma_{i}}+\delta_{\sigma_{i+1}\bar\sigma_{i+1}} \right)
\right] \, ,
\nonumber
\label{WB}
\end{eqnarray}
where we consider the ferromagnetic case ($J < 0$) throughout this paper.

The variational lower bound for the partition function per row is the 
maximum of the Rayleigh ratio
\begin{equation}
\lambda =
\frac{\sum\limits_{[ \bar\sigma ] , [ \sigma ]}\Phi[ \bar\sigma ]
{\cal T}[ \bar\sigma | \sigma ] \Psi[ \sigma ]}{
\sum\limits_{[ \bar\sigma ] , [ \sigma ]}
\Phi[ \bar\sigma] \Psi[ \sigma ]}
\equiv  \frac{\langle \Phi |{ \cal T} | \Psi \rangle}{\langle \Phi | \Psi \rangle} \, ,
\label{Rr}
\end{equation}
where $\Phi[ \bar\sigma ]$ and $\Psi[ \sigma ]$ are arbitrary variational
states. Since the transfer matrix ${\cal T}$ in Eq.~(\ref{T2D}) is symmetric,
we assume $\Phi[ \sigma ] = \Psi[ \sigma ]$ in the following.

\begin{figure}[!ht]
\includegraphics[width=70mm,clip]{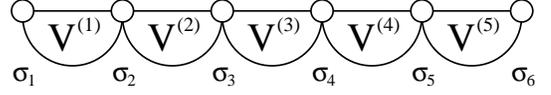}
\caption{Graphical expression of the variational state $\Psi[ \sigma ]$ in
Eq.~(\ref{V2}).}
\label{VV2D}
\end{figure}

\begin{figure}[!ht]
\includegraphics[width=70mm,clip]{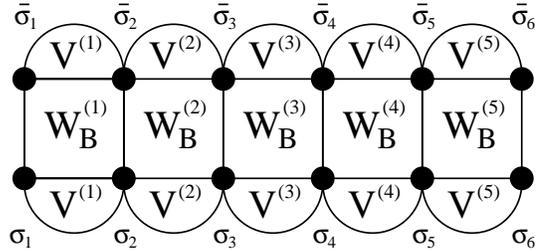}
\caption{Graphical expression of $\langle \Psi | {\cal T} | \Psi \rangle$ in 
Eq.~(\ref{2Dtmvwv}). We have used black circles for the spins whose configuration
sum is taken.}
\label{VTV}
\end{figure}

TPVA consists of local approximations~\cite{TPVA1,TPVA2} which restrict the
form of the variational state $\Psi[ \sigma ]$ into a uniform product of local
factors
\begin{equation}
\Psi[ \sigma ]
= \prod\limits_{i=1}^{2N-1}V^{(i)}\{ \sigma \}
= \prod\limits_{i=1}^{2N-1}V( \sigma_{i}^{~} \sigma_{i+1}^{~} )  \, ,
\label{V2}
\end{equation}
where there are only $q^2_{~}$ variational parameters. Figure~\ref{VV2D}
graphically represents $\Psi[ \sigma ]$ when $2N = 6$. 
A profit of writing the variational state in the product form is that the norm
of the variational state also has the local product structure
\begin{eqnarray}
\langle \Psi | \Psi \rangle
&=& \sum_{[ \sigma ]}^{~} \prod_{i = 1}^{2N - 1}
\left( V_{~}^{(i)}\{ \sigma \} \right)^2_{~} \nonumber\\
&=& \sum_{[ \sigma ]}^{~} \prod_{i = 1}^{2N - 1}
\left( V( \sigma_i^{~} \sigma_{i+1}^{~} ) \right)^2_{~},
\end{eqnarray}
which is nothing but a partition function of a 1D lattice model whose
local Boltzmann weight is
$\left( V( \sigma_i^{~} \sigma_{i+1}^{~} ) \right)^2_{~}$.
In the same manner, the numerator of Eq.~(\ref{Rr}) is written as
\begin{equation}
\langle \Psi | {\cal T} | \Psi \rangle
= \sum_{[ \bar\sigma ] [ \sigma ]}^{~} \prod_{i = 1}^{2N - 1}
V_{~}^{(i)}\{ \bar\sigma \} W_{\rm B}^{(i)}\{ \bar\sigma | \sigma \}
V_{~}^{(i)}\{ \sigma \} \, ,
\label{2Dtmvwv}
\end{equation}
which is also a partition function of an effective 2-leg ladder.
As we have graphically represented the variational state $\Psi[ \sigma ]$ in 
Fig.~\ref{VV2D}, let us also express
$\langle \Psi | {\cal T} | \Psi \rangle$ graphically in Fig.~\ref{VTV}.

\begin{figure}[!ht]
\includegraphics[width=70mm,clip]{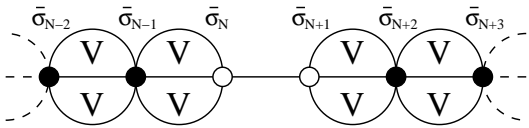}
\caption{The factor $A^{(N)}$ in Eq.~(\ref{eqa}) is constructed
by joining two $\Psi[\sigma]$s and taking spin configuration sum 
over all spins $[\sigma]$ (the black circles) except for the two
central ones $\{\bar\sigma\}=(\bar\sigma_{N},\,\bar\sigma_{N+1})$
(the white circles).}
\label{M_A}
\end{figure}

\begin{figure}[!ht]
\includegraphics[width=70mm,clip]{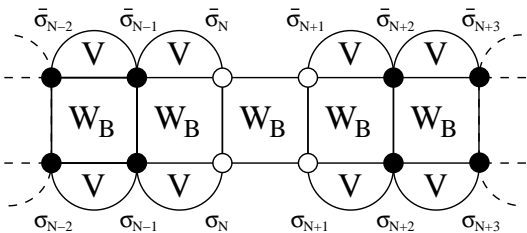}
\caption{Graphical representation of $B^{(N)}_{~}$ in Eq.~(\ref{eqb}).}
\label{M_B}
\end{figure}

With the use of the variational state thus defined, the variational problem
in Eq.~(\ref{Rr}) is the same as those used by Villani~\cite{Villani,BAXbook}. 
Our aim is to obtain the best local factor
$V\{\sigma\}$ numerically. There are several ways to maximize
$\lambda_{\rm var}^{~}$ in Eq.~(\ref{Rr}), under the condition that the lattice
size $2N$ is sufficiently large~\cite{TPVA1,TPVA2}. Keeping the extension
to three dimensions in our mind, what we consider here is to take the
variations of $\lambda_{\rm var}^{~}$ with respect to each local factor
\begin{equation}
\frac{\delta\lambda_{\rm var}^{~}}{\delta\Psi}
\equiv
\sum\limits_{i}\frac{\delta\lambda_{\rm var}^{~}}{\delta V^{(i)}_{~}} \, .
\label{varl}
\end{equation}
When the system size $2N$ is large
enough, it is sufficient to consider the variation
with respect to the local change
\begin{equation}
V^{(N)}_{~} \rightarrow V^{(N)}_{~} +\delta V^{(N)}_{~}
\end{equation}
at the center of the spin row, since we have treated the uniform variational 
state and since the boundary effect is negligible. After a short 
calculation from the (local) extremal condition~\cite{TPVA1,TPVA2}
\begin{equation}
\frac{\delta\lambda}{\delta V^{(N)}_{~}} = 0,
\end{equation}
we obtain an eigenvalue problem
\begin{equation}
\sum\limits_{\{\sigma\}}
\frac{B^{(N)}_{~}\{ \bar\sigma | \sigma \}}{A^{(N)}_{~}\{ \bar\sigma \}}\,
V^{(N)}_{~}\{ \sigma \} = \lambda\,  V^{(N)}_{~}\{ \bar\sigma \} \,
\label{selfc}
\end{equation}
for the local factor $V^{(N)}$. The new factor $A^{(N)}_{~}\{ \bar\sigma \}$
is constructed as
\begin{eqnarray}
\!\!\!\!\!\!\!\!
\label{eqa}
A^{(N)}_{~}\{ \bar\sigma \}
&=& A\{ \bar\sigma_N^{~} \bar\sigma_{N+1}^{~} \} \\
&=&
\sum_{ \bar\sigma_1^{~} \cdots \bar\sigma_{N-1}^{~}
\bar\sigma_{N+2}^{~} \cdots \bar\sigma_{2N}^{~}}^{~}
\prod_{i \neq N}^{~} \left( V^{(i)}\{ \bar\sigma \} \right)^2_{~} \, ,
\nonumber
\end{eqnarray}
whose graphical representation is shown in Fig.~\ref{M_A}.
The matrix $B^{(N)}_{~}$ is defined in the same manner
\begin{eqnarray}
\!\!\!\!\!\!\!\!
B^{(N)}_{~}\{ \bar\sigma | \sigma \} =
W_{\rm B}^{(N)}\{ \bar\sigma | \sigma \} \sum_{[ \bar\sigma ] [ \sigma ]}
^{i \neq N, N+1}
\prod_{ i \neq N }^{~} & V^{(i)}\{ \bar\sigma \}& \nonumber\\
\times  W_{\rm B}^{(i)}\{ \bar\sigma | \sigma \}  &V^{(i)}\{\sigma\}& \, ,
\label{eqb}
\end{eqnarray}
where the spin configuration sum is taken over all black circles
in Fig.~\ref{M_B}.

Since both $A^{(N)}_{~}$ and $B^{(N)}_{~}$ are constructed from
the local factor $V$, the eigenvalue relation Eq.~(\ref{selfc}) should be solved
self-consistently. Thus Eq.~(\ref{selfc}) is a kind of the self-consistent 
equation. A realistic outline how to solve the self-consistent equation
is as follows.
\begin{itemize}
\item{} Start the calculation by setting (arbitrary) $q^2_{~}$ numbers of
initial values for the local factor $V( \sigma, \sigma' )$.
\item{} Calculate $A^{(N)}_{~}$ and $B^{(N)}_{~}$ from Eqs.~(\ref{eqa})
and (\ref{eqb}), respectively,
for sufficient large system size $2N$.
\item{} Substitute $A^{(N)}_{~}$, $B^{(N)}_{~}$, and $V^{(N)}_{~}$ to the
left hand side of Eq.~(\ref{selfc}). Obtain the right hand side by
\begin{equation}
V'\{\bar\sigma\}=\sum\limits_{\{\sigma\}}\frac{B^{N}_{~}\{\bar\sigma|\sigma\}}
{A^{N}_{~}\{\bar\sigma\}}V^{N}_{~}\{\sigma\}
\end{equation}
and normalize it
\begin{equation}
V''\{\sigma\}=\frac{V'\{\sigma\}}{\sqrt{\sum\limits_{\{\sigma'\}}
{\left(V'\{\sigma'\}\right)}^2}}.
\end{equation}
\item{} Create a linear combination $V_{\rm new}^{~} = V + \varepsilon V''$
where $\varepsilon$ is a small parameter of the order of $0.1$, and regard it as
an improved local factor. After normalizing $V_{\rm new}^{~}$ go to the second 
step and repeat the calculation till $V$ reaches its (local) fixed point.
\end{itemize}
The small parameter $\varepsilon$ is introduced in order to stabilize the
convergence of the iterative calculation. For statistical models that exhibit a
phase transition, the self-consistent equation has several stable solutions near 
the transition temperature. They correspond to the disordered state and to each 
ordered state. In such a case, one can `target' a desired phase
just by imposing a very small symmetry-breaking field or by setting the 
initial local factor $V( \sigma, \sigma' )$ appropriately.

The main advantage of the above algorithm is that no a priori
ansatz is necessary for setting up the variational parameters. 

\section{Extension to three dimensions}

\begin{figure}[!ht]
\includegraphics[width=80mm]{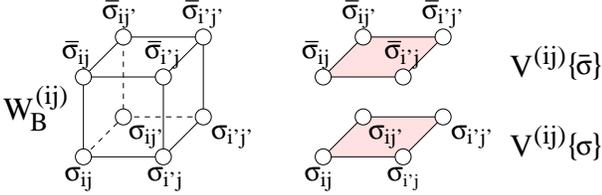}
\caption{The IRF type local Boltzmann weight
 $W_{\rm B}^{(ij)}\{ \bar\sigma | \sigma \}$
of the $q$-state Potts models and the variational factors
$V_{~}^{(ij)}\{ \bar\sigma \}$ and $V_{~}^{(ij)}\{ \sigma \}$.
The $q$-state variables $\sigma = 0, 1, \dots, q - 1$
are located at the edges of the cube. We use the notation $\{ \bar{\sigma} \}$
and $\{ {\sigma} \}$ for the upper and the lower horizontal plaquettes.}
\label{fig1}
\end{figure}

It is easy to generalize both the variational relation (Eq.~(\ref{Rr})) and the 
construction of the variational state in the product form (Eq.~(\ref{V2})) to 
3D models. We can increase the space dimension by replacing the 
row-spin $[ \sigma ]$ in Eq.~(\ref{spin}) to a `layer spin'
\begin{equation}
[ \sigma ] =
\left(
\begin{array}{cccccc}
\sigma_{1~ \, 1}^{~} & \cdots & \sigma_{1~ \, N}^{~} & \sigma_{1~ \, N+1}^{~} &
\cdots & \sigma_{1~ \, 2N}^{~}\\
\vdots       & \ddots & \vdots  & \vdots   & \ddots & \vdots \\
\sigma_{N~~ \, 1}^{~} & \cdots & \sigma_{N~~ \, N}^{~} & \sigma_{N~~ \, N+1}^{~} &
\cdots & \sigma_{N~~ \, 2N}^{~}\\
\sigma_{N+1 \, 1}^{~} & \cdots & \sigma_{N+1 \, N}^{~} & \sigma_{N+1 \, N+1}^{~} &
\cdots & \sigma_{N+1 \, 2N}^{~}\\
\vdots       & \ddots & \vdots  & \vdots & \ddots & \vdots \\
\sigma_{2N~ \, 1}^{~} & \cdots & \sigma_{2N~ \, N}^{~} & \sigma_{2N~ \, N+1}^{~} &
\cdots & \sigma_{2N~ \, 2N}^{~}\\
\end{array}
\right) \, .
\end{equation}
Now the system we are considering is an infinitely large 3D object of size
$2N\times2N\times\infty$. As before, we assume that the system size $2N$ is 
sufficiently large to investigate the bulk limit.
The 3D generalization of the row-to-row transfer matrix in Eq.~(\ref{T2D}) is
a layer-to-layer transfer matrix. For the 3D $q$-state Potts model, the
layer-to-layer transfer matrix is given by
\begin{equation}
{\cal T}[ \bar\sigma | \sigma ] = \prod_{i=1}^{(2N-1)} \prod_{j=1}^{(2N-1)}
W_{\rm B}^{(ij)}\{ \bar\sigma | \sigma \} \, ,
\end{equation}
where the IRF-type local Boltzmann weight is written as
\begin{eqnarray}
&~& W_{\rm B}^{(ij)}\{ \bar\sigma | \sigma \} = W_{\rm B}^{~}\left\{
{\bar\sigma_{ij}\ \bar\sigma_{i'j}\ \bar\sigma_{i'j'}\ \bar\sigma_{ij'}}
\atop{\sigma_{ij}\ \sigma_{i'j}\ \sigma_{i'j'}\ \sigma_{ij'}}
\right\}\\
\nonumber
=\exp\hspace{-0.2cm} & \biggl[ \frac{-J}{4k_{\rm B}T} & \hspace{-0.2cm}\left(
\delta_{\sigma_{ij}\sigma_{i'j}}+\delta_{\sigma_{i'j}\sigma_{i'j'}}+
\delta_{\sigma_{i'j'}\sigma_{ij'}}+\delta_{\sigma_{ij'}\sigma_{ij}}
\right. \\
& & \nonumber
\hspace{-0.2cm}+\delta_{\bar\sigma_{ij}\bar\sigma_{i'j}}+
\delta_{\bar\sigma_{i'j}\bar\sigma_{i'j'}}+
\delta_{\bar\sigma_{i'j'}\bar\sigma_{ij'}}+
\delta_{\bar\sigma_{ij'}\bar\sigma_{ij}} \\
& & \left.
\hspace{-0.2cm}+\delta_{\sigma_{ij}\bar\sigma_{ij}}+
\delta_{\sigma_{i'j}\bar\sigma_{i'j}}+
\delta_{\sigma_{i'j'}\bar\sigma_{i'j'}}+
\delta_{\sigma_{ij'}\bar\sigma_{ij'}}
\right) \biggr] \, .
\nonumber
\label{WB3D}
\end{eqnarray}
We have used the notation $i' = i + 1$ and $j' = j + 1$, and have represented
the plaquette spins as $\{ \sigma \}$. (See Fig.~\ref{fig1}.)

The 2D generalization of the variational state in Eq.~(\ref{V2}) can be obtained in
the same manner
\begin{eqnarray}
\Psi[ \sigma ]
&=& \prod_{i=1}^{(2N-1)} \prod_{j=1}^{(2N-1)}
V^{(ij)}_{~}\left\{ \sigma \right\} \nonumber\\
&=& \prod_{i=1}^{(2N-1)} \prod_{j=1}^{(2N-1)}
V( \sigma_{ij}\ \sigma_{i'j}\ \sigma_{i'j'}\ \sigma_{ij'} ) \, .
\label{V2_3D}
\end{eqnarray}
There are $q^4$ variational parameters in the local factor
$V^{(ij)}_{~}$. We assume that the factor $V^{(ij)}_{~}$ is positionally 
independent and the variational state is uniform. The local factor at the center
of the system is $V^{(NN)}_{~}$.

The way how to optimize the local factor $V^{(ij)}_{~}$, so that it maximizes
the Rayleigh ratio $\lambda_{\rm var}^{~}$ in Eq.~(\ref{Rr}), is in principle
the same as that for 2D systems. The denominator 
\begin{equation}
\langle \Psi | \Psi \rangle = \sum_{[ \sigma ]}^{~} \prod_{i=1}^{(2N-1)} 
\prod_{j=1}^{(2N-1)}
\left( V^{(ij)}_{~}\{ \sigma \} \right)^2_{~}
\end{equation}
is nothing but a partition function of a 2D lattice model
whose local Boltzmann weight is equal to $\left( V^{(ij)}_{~} \right)^2_{~}$,
and the numerator $\langle \Psi | {\cal T} | \Psi \rangle$
is that of a two-layer 2D lattice model
\begin{equation}
\sum_{[ \bar\sigma ] [ \sigma ]}^{~} \prod_{i=1}^{(2N-1)} \prod_{j=1}^{(2N-1)}
V^{(ij)}_{~}\{ \bar\sigma \}
W_{\rm B}^{(ij)}\{ \bar\sigma | \sigma \}
V^{(ij)}_{~}\{ \sigma \} \, .
\end{equation}
Since the numerator and the denominator are the partition functions of
effective 2D lattice models one can calculate both of them using the corner 
transfer matrix renormalization group (CTMRG), which is a variant of DMRG
applied to 2D lattice models~\cite{CTMRG}. As a byproduct of CTMRG,
the factor $A^{(NN)}_{~}\{ \sigma \}$ and the matrix
$B^{(NN)}_{~}\{ \bar\sigma | \sigma \}$ can be calculated~\cite{abctmrg}.
Also, the variational free energy per site
$\langle F \rangle$ can be obtained from CTMRG. (Numerical details
are reported in Ref.~\cite{KW_ON,TPVA1,TPVA2}.)

After we obtain the optimized variational factor $V^{(NN)}_{~}\{ \sigma \}$,
the internal energy $E$ and the magnetization $M$ can be calculated from
$A^{(NN)}_{~}\{ \sigma \}$ and $B^{(NN)}_{~}\{ \bar\sigma | \sigma \}$ that are
created from the optimized variational factor $V^{(NN)}$.
The internal energy $E$ per site is equivalent to
\begin{eqnarray}
E =
&-& J\, \delta( \sigma_{N \, N}^{~},\, \sigma_{N+1\, N}^{~} )
- J\, \delta( \sigma_{N \, N}^{~},\, \sigma_{N \, N+1}^{~} ) \nonumber\\
&-& J\, \delta( \sigma_{N \, N}^{~},\, \bar\sigma_{N \, N}^{~} )
\label{ie}
\end{eqnarray}
and its statistical average is obtained as follows
\begin{equation}
\langle E \rangle = \frac{\displaystyle \sum_{\{ \bar\sigma \} \{ \sigma \}}^{~} 
\, E \, V\{ \bar\sigma \} B\{ \bar\sigma | \sigma \} V\{ \sigma \} }
{\displaystyle \sum_{\{ \sigma \}}^{~} \, V\{\sigma \}A\{ \sigma \}V\{\sigma\}}\, ,
\end{equation}
where we have dropped the superscript $(NN)$ from $V$, $A$, and $B$ just for
simplicity. The magnetization $\langle M \rangle$ of the $q$-state Potts 
model can be calculated from the spin expectation value
\begin{equation}
\langle \delta( \sigma, \, 0 ) \rangle =
\frac{\displaystyle \sum_{\{ \bar\sigma \} \{ \sigma \}}^{~} \,
\delta( \sigma_{NN}^{~} , \, 0 ) \, V\{ \bar\sigma \}B\{ \bar\sigma | \sigma \}
V\{ \sigma \} }{\displaystyle  \sum_{\{ \sigma \}}^{~} \, V\{\sigma \}
A\{ \sigma \}V\{\sigma\}}
\end{equation}
together with the definition of the order parameter
\begin{equation}
\langle M \rangle = \frac{q \langle \delta( \sigma, \, 0 ) \rangle - 1}{q - 1} \, .
\label{mg}
\end{equation}

\section{Numerical results}

\begin{figure}[!ht]
\includegraphics[width=85mm,clip]{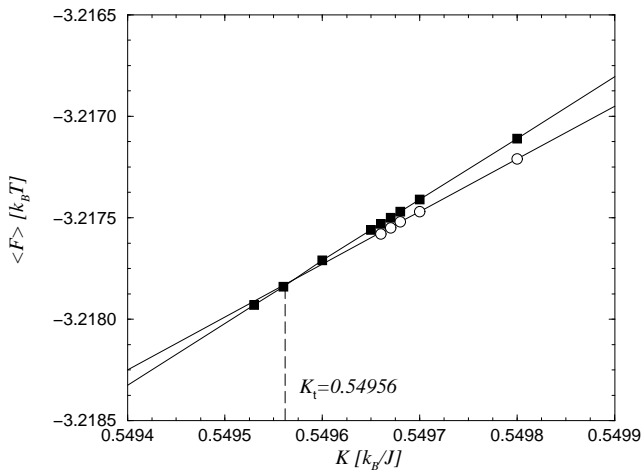}
\caption{The free energy per site $\langle F \rangle$ of the $q = 3$ Potts
model with respect to the inverse temperature $K = 1 / T$.}
\label{fig3a}
\end{figure}
\begin{figure}[!ht]
\includegraphics[width=85mm,clip]{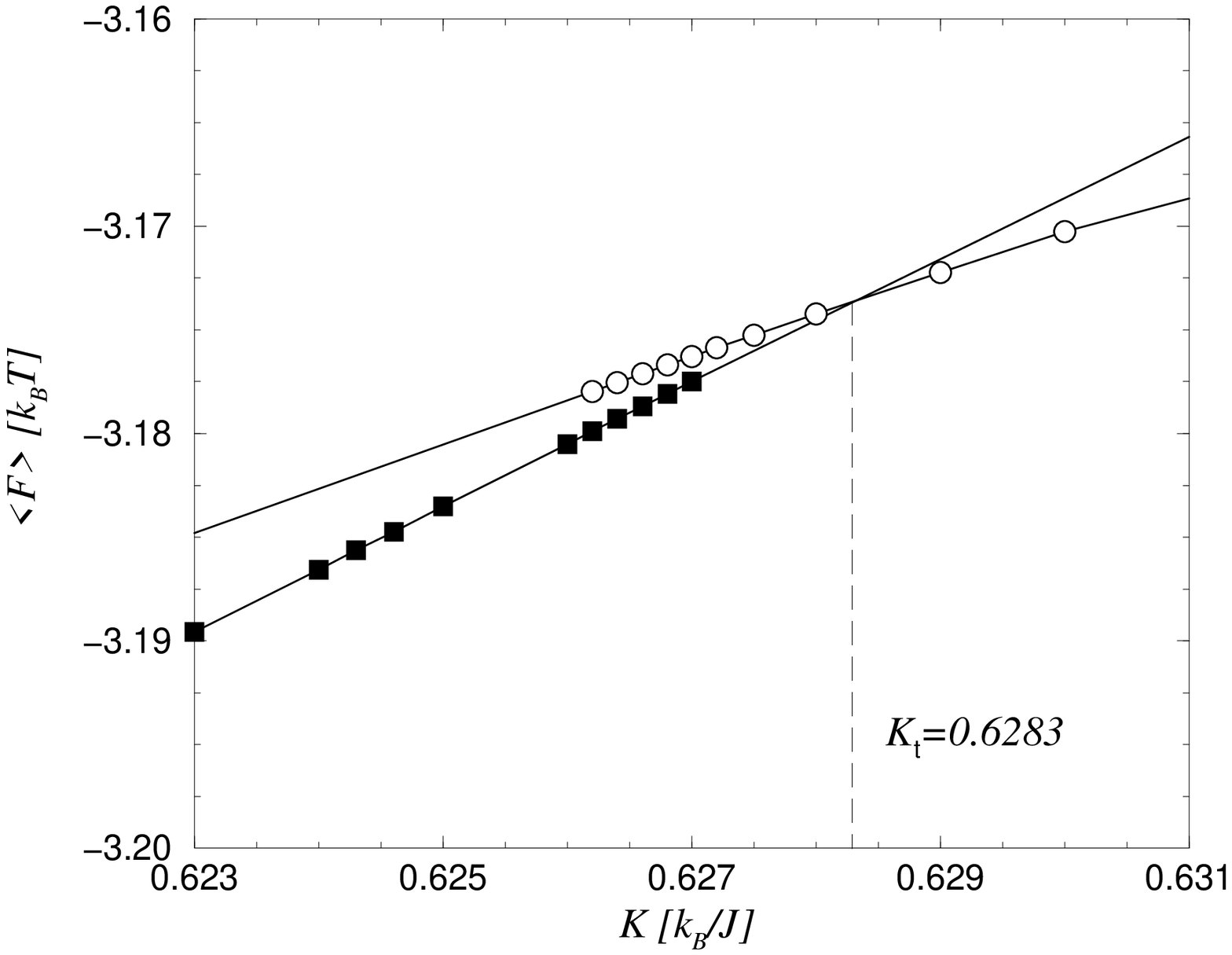}
\caption{The free energy per site $\langle F \rangle$ of the $q = 4$ Potts model.}
\label{fig3b}
\end{figure}
\begin{figure}[!ht]
\includegraphics[width=85mm,clip]{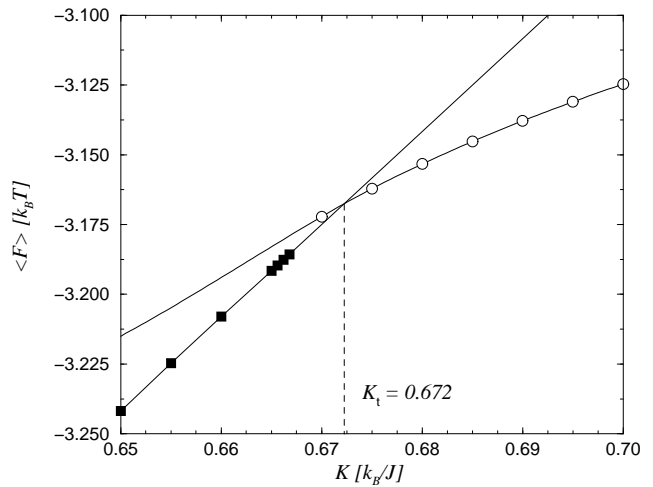}
\caption{The free energy per site $\langle F \rangle$ of the $q = 5$ Potts model.}
\label{fig3c}
\end{figure}

We calculate the latent heat of the 3D $q = 3$, $4$, and $5$ Potts
models, using the internal energy expectation values $\langle E \rangle$ for
both ordered and disordered phases. Hereafter we set $k_{\rm B}^{~}
= \mu_{\rm B}^{~} = 1$ and only treat the ferromagnetic case $J = -1$.
The convergence control parameter in the self-consistent calculation
is chosen as $\varepsilon = 0.1$. When we obtain the variational state
for the ordered phase, we impose a small symmetry breaking field 
($\sim$~magnetic field) to the system during first several 
iterations, and after that we switch it off. For the CTMRG calculations,
we kept block spin states $m$ up to the value of 20~\cite{mvar}, which
is sufficiently large to obtain the thermodynamic functions shown bellow.
All thermodynamic functions converged after 500 iterations at most
even in the close vicinity of the transition point.

First we determined the transition temperature from the calculated
free energy per site $\langle F \rangle$ with respect to the
inverse temperature $K \equiv 1 / T$.
Since the $q = 3 \sim 5$ Potts models exhibit the first-order phase 
transitions, in a close vicinity of the transition point $K_{\rm t}$
there are two minima in the free energy $F$; one corresponds to
the disordered phase and the other to the ordered one. It is possible
to detect both of them by way of solving the self-consistent equation
starting from different initial conditions for local factors. (When
the barrier between the minima is low, one of the two phases is often
accidentally chosen by numerical round-off errors.)
Figs.~\ref{fig3a}, \ref{fig3b} and \ref{fig3c}, respectively, show
the calculated free energy per site $\langle F \rangle$ for $q = 3$,
$4$, and $5$ cases. The black squares and the white circles represent
 $\langle F \rangle$ for disordered and ordered phases, respectively, where 
the point of intersection of these two curves results the transition point 
$K_{\rm t}^{~}$. The free energy curves were drawn by the least-square
fitting of plotted data to polynomials. The results are,
$K_{\rm t}^{[q=3]} = 0.5496$ for $q = 3$,
$K_{\rm t}^{[q=4]} = 0.6283$ for $q = 4$, and
$K_{\rm t}^{[q=5]} = 0.672$ for $q = 5$.
For the case $q = 3$ the most reliable Monte Carlo result (as far as we know) is
$K_{\rm t}^{\rm MC} = 0.550565 \pm0.000010$~\cite{MC}, and
thus $K_{\rm t}^{[q=3]}$ calculated by TPVA is only 0.18\% lower than
$K_{\rm t}^{\rm MC}$.

\begin{figure}[!ht]
\includegraphics[width=85mm,clip]{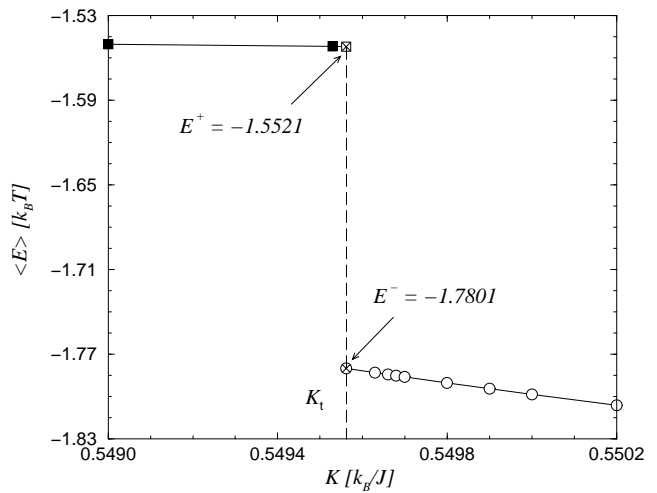}
\caption{The energy per site $\langle E \rangle$ with respect to $K$ for the $q = 3$
Potts model.}
\label{fig4a}
\end{figure}
\begin{figure}[!ht]
\includegraphics[width=85mm,clip]{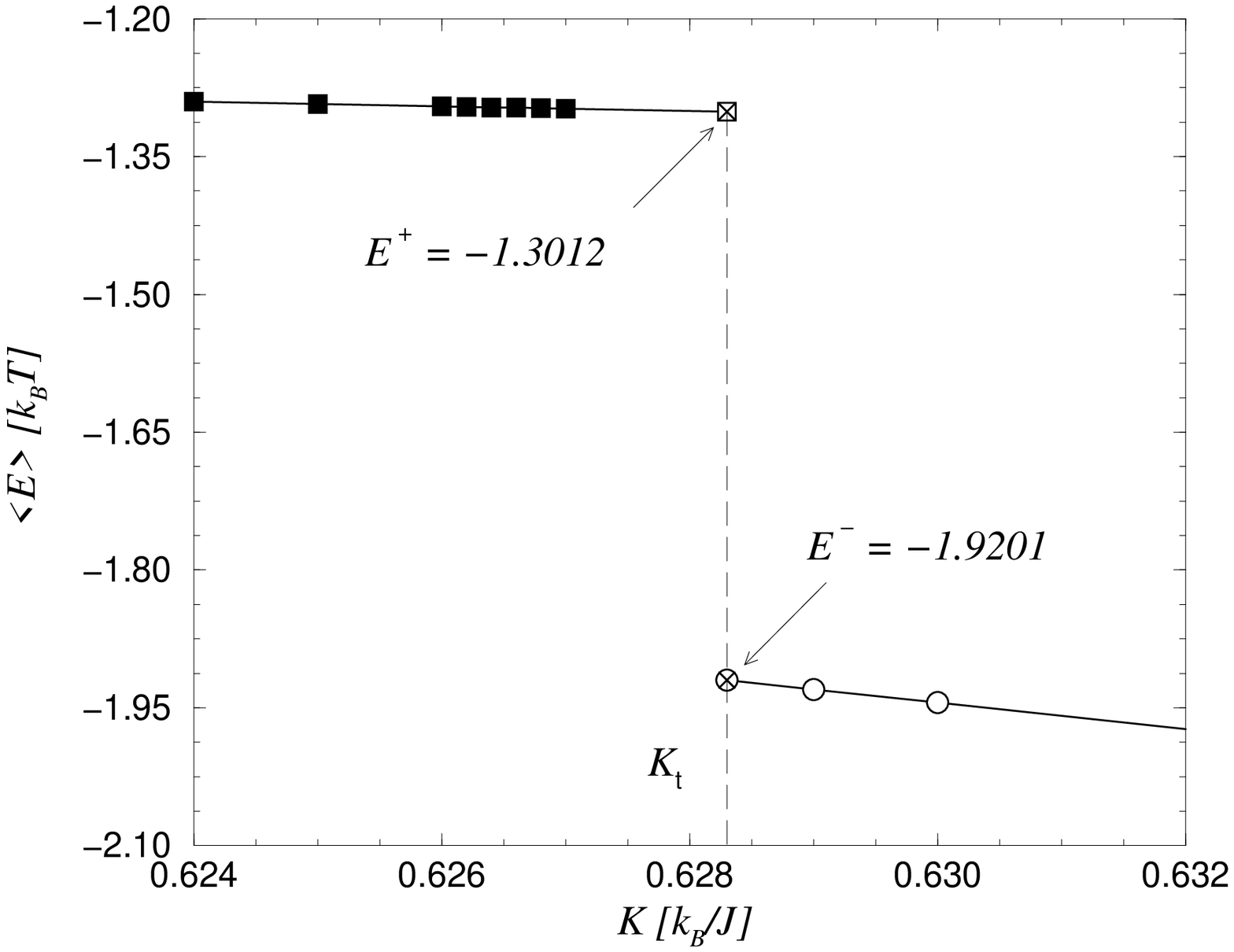}
\caption{The energy per site $\langle E \rangle$ with respect to $K$ when $q = 4$.}
\label{fig4b}
\end{figure}
\begin{figure}[!ht]
\includegraphics[width=85mm,clip]{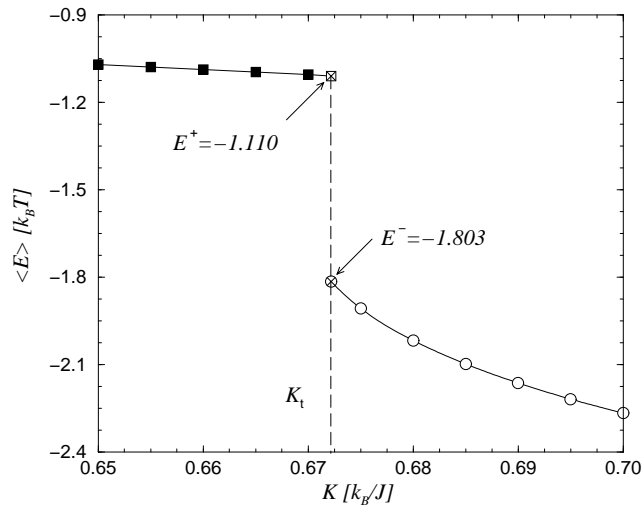}
\caption{The energy per site $\langle E \rangle$ with respect to $K$ when $q = 5$.}
\label{fig4c}
\end{figure}
In Figs.~\ref{fig4a}, \ref{fig4b}, and \ref{fig4c}, we have plotted the
internal energies per site $\langle E \rangle$ as functions of
$K \equiv 1 / T$. The latent heat is the energy difference
\begin{equation}
Q = E_{~}^+ - E_{~}^-
\label{lheat}
\end{equation}
between the ordered and disordered phases. As before, we have applied the 
least-square fittings to interpolate (or extrapolate) the calculated data
towards $E_{~}^+$ and 
$E_{~}^-$ at the determined transition point $K_{\rm t}^{[q=3,4,5]}$. These
energies $E_{~}^+$ and $E_{~}^-$, respectively, are denoted by the cross 
symbols inside the squares and the circles in Figs.~\ref{fig4a}, \ref{fig4b}, 
and \ref{fig4c}. The results are
$Q_{~}^{[q = 3]} = 0.228$,
$Q_{~}^{[q = 4]} = 0.619$, and
$Q_{~}^{[q = 5]} = 0.693$. For the case $q = 3$, $Q_{~}^{[q = 3]}$ is
41\% larger than a Monte Carlo result
$Q_{\rm MC}^{[q = 3]} = 0.16160 \pm 0.00047$.~\cite{MC}

\begin{figure}[!ht]
\includegraphics[width=85mm,clip]{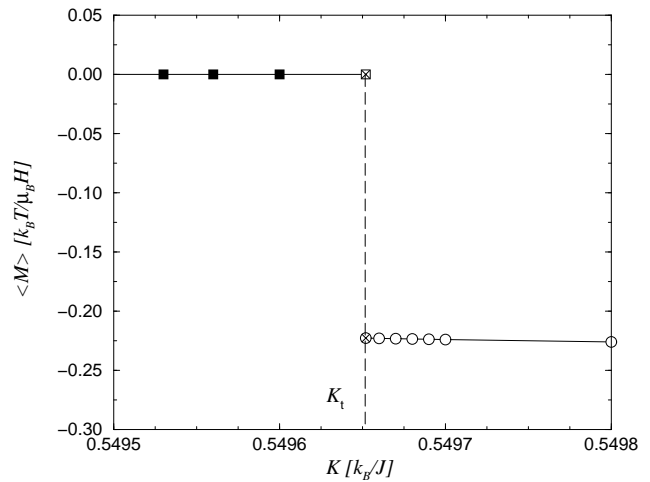}
\caption{The magnetization $\langle M\rangle$ with respect to the inverse temperature
$K$ for $q = 3$.}
\label{fig2a}
\end{figure}

\begin{figure}[!ht]
\includegraphics[width=85mm,clip]{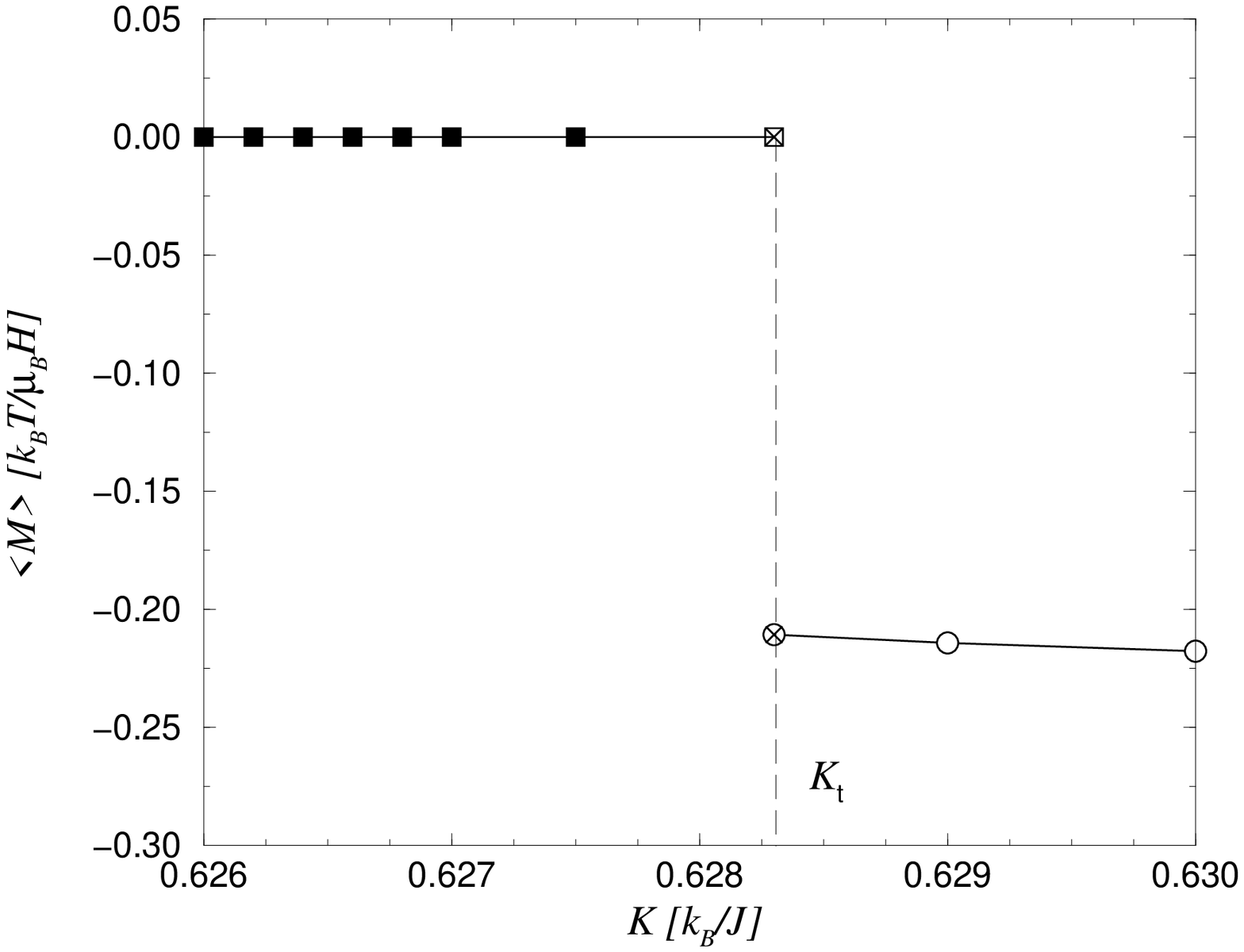}
\caption{The magnetization $\langle M\rangle$ with respect to $K$ for
$q = 4$.}
\label{fig2b}
\end{figure}

\begin{figure}[!ht]
\includegraphics[width=85mm,clip]{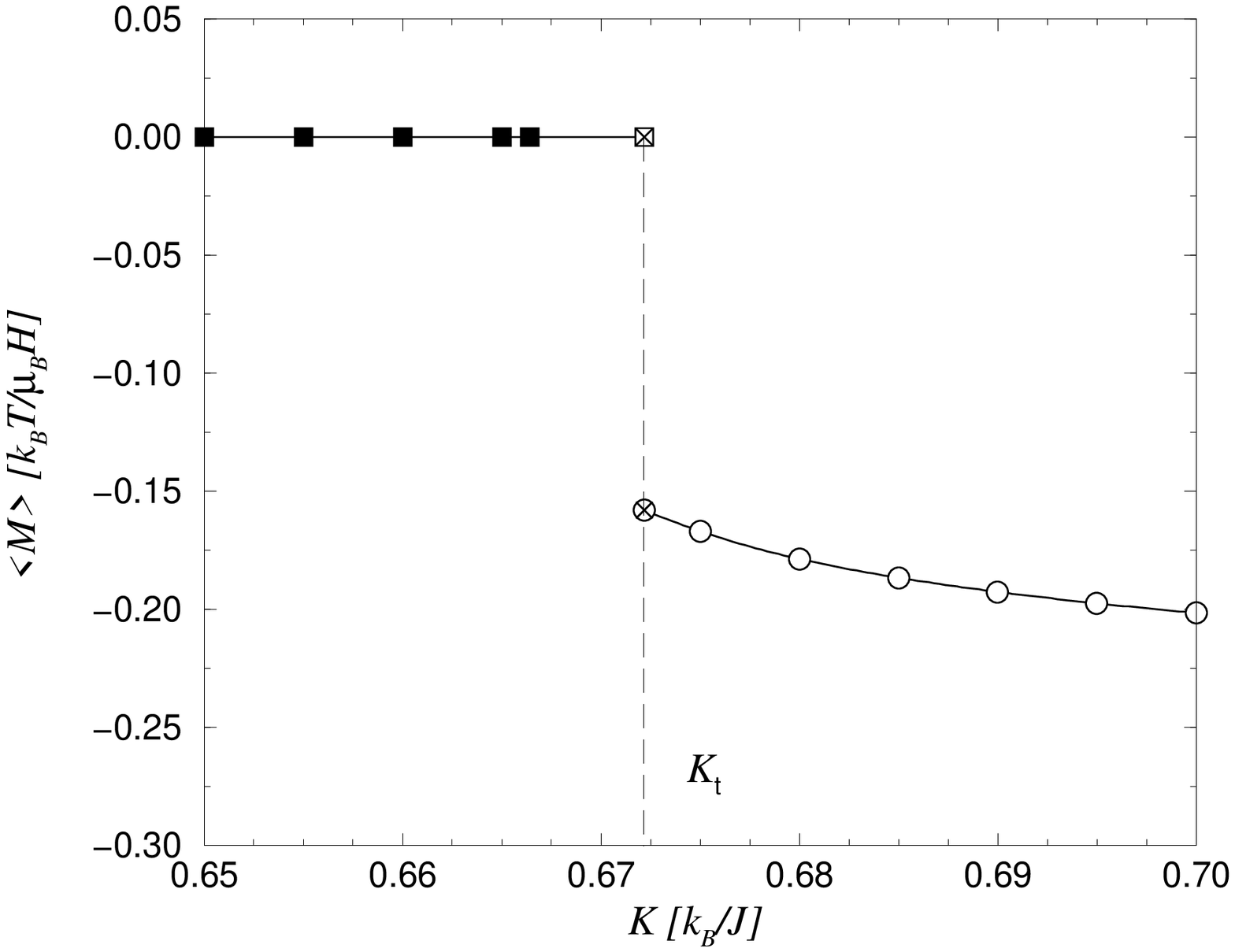}
\caption{The magnetization $\langle M\rangle$ with respect to $K$ for
$q = 5$.}
\label{fig2c}
\end{figure}

We finally show the calculated spontaneous magnetization $\langle M \rangle$
 in Figs.~\ref{fig2a}, \ref{fig2b}, and \ref{fig2c}.
All numerical results thus obtained are summarized in Table~I.

\begin{table}[!ht]
\caption {The numerically obtained transition points $K_{\rm t}$ and the
latent heats $Q$ by TPVA for the 3D ferromagnetic $q$=3, 4, and 5 state
Potts models. The values of the $m$-state block spins are given in the 
second column.}
\label{t1}
\begin{ruledtabular}
\begin{tabular*}{\hsize}{
l@{\extracolsep{0ptplus1fil}}
r@{\extracolsep{0ptplus1fil}}
c@{\extracolsep{0ptplus1fil}}c}
$q$ & $m$ & $K_{\rm t}$ & $Q$ \\
\colrule
3 & 20 & 0.5496 & 0.228 \\
4 & 20 & 0.6283 & 0.619 \\
5 & 5 & 0.672\phantom{2} & 0.693 \\
\end{tabular*}
\end{ruledtabular}
\end{table}

\section{Conclusions}

Recently proposed self-consistent method for 3D classical systems, the TPVA,
has been applied to $q$=3, 4, and 5 state Potts models on the simple cubic lattice.
Thermodynamic functions such as the free energy, the internal energy, and the
spontaneous magnetizations are calculated. The numerical algorithm for solving 
the self-consistent equation is stable at any temperature, if the convergence 
control parameter $\varepsilon$ is chosen to be equal or smaller than $0.1$.

\begin{acknowledgments}
The authors thank to Y.~Hieida, K.~Okunishi, N.~Maeshima, and Y.~Akutsu for
discussions about variational formulations.
This work has been partially supported by the Slovak Grant Agencies, VEGA
No.~2/7201/21 and Grant-in-Aid for Scientific Research from
Ministry of Education, Science, Sports and Culture (No.~09640462 and
No.~11640376). A.G. is supported by Japan Society for the Promotion of
Science (P01192). The numerical calculations were performed by
Compaq Fortran on the HPC-Alpha UP21264 Linux workstation.
\end{acknowledgments}


\begin{thebibliography}{99}
\bibitem{abctmrg} One can calculate the factor $A^{(NN)}_{~}\{ \sigma \}$ and 
the matrix $B^{(NN)}_{~}\{ \bar\sigma | \sigma \}$ by combining the 
renormalized corner transfer matrix and the renormalized half-row transfer matrix.
\end{thebibliography}
\end{document}